\documentclass[twocolumn,showpacs,amsmath,amssymb,prb]{revtex4}
\usepackage{graphicx}

\begin{document}
\title{Thermal rectifying effect in two dimensional anharmonic lattices}

\author{Jinghua Lan}
\affiliation{Department of Physics and Centre for Computational
Science and Engineering, National University of Singapore,
Singapore 117542}
 \author{Baowen Li}
 \email{phylibw@nus.edu.sg}
 \affiliation{Department of Physics and Centre for Computational
Science and Engineering, National University of Singapore,
Singapore 117542} \affiliation{Laboratory of Modern Acoustics and
Institute of Acoustics, Nanjing University, 210093, PR China}
\affiliation{NUS Graduate School for Integrative Sciences and
Engineering, Singapore 117597, Republic of Singapore}
\date{1 January 2007, Published in Phys. Rev. B 74, 214305 (2006).}

\begin{abstract}
We study thermal rectifying effect in two dimensional (2D) systems
consisting of the Frenkel Kontorva (FK) lattice and the
Fermi-Pasta-Ulam (FPU) lattice. It is found that the rectifying
effect is related to the asymmetrical interface thermal
resistance. The rectifying efficiency is typically about two
orders of magnitude which is large enough to be observed in
experiment. The dependence of rectifying efficiency on the
temperature and temperature gradient is studied. The underlying
mechanism is found to be the match and mismatch of the spectra of
lattice vibration in two parts.
\end{abstract}
\pacs{67.40.Pm, 63.20.Ry, 66.70.+f, 44.10.+i} \maketitle

\section{Introduction}
Heat conduction in low dimensional systems has attracted
increasing attention \cite{Peierls55, Casati84, Prosen92,
Kaburaki93, Lepri97, Lepri98, SLepri98, Lepri03,
Hu98,Fillipov,Hu00, Bonetto00, Giardin00, Gendelman00, Prosen00,
Aoki,Li02, Alonso02, Li03,Pereverzev03, BLi03, Li04,Li06} in
recent years. After two decades analytic and numerical studies in
1D model, much progress has been achieved. On the one hand, the
study has enriched our understanding about the underlying physical
mechanism. On the other hand, the study has made it possible to
seek the practical application of heat control and management.
Indeed, in 2002, Terraneo et al \cite{Terraneo02} proposed a
thermal device which can rectify the heat current through it when
reversing the temperature gradient. The model proposed by Terraneo
et al is a 1D anharmonic lattice consisting of three segments with
the Morse on-site potential of different parameters. As the first
attempt of controlling heat current, the ratio of the thermal
current changes is less than two. More recently, Li et al.
\cite{BLi04} construct a thermal diode model in which two
Frenkel-Kontoroval (FK) chains with different nonlinear strengths
are connected by a harmonic spring.
 The most successful improvement of the model by
Li et al \cite{BLi04} lies in three facts: First, the
configuration is more simple, it consists of only two different
segments; Second, the ratio of heat current from two different
directions is increased drastically about 100 times; Third, the
underlying mechanism of thermal rectifying effect is explained by
illustrating the phonon bands of the particles in different
segments. Following Li et al's work, we further improve the
rectifying efficiency from 100 to 2000 by substituting the weak FK
chain with a Fermi-Pasta-Ulam (FPU) chain \cite{Li05}. In
addition, we find that the rectifying effect (asymmetric heat
flow) is closely related to asymmetric interface thermal
resistance (also called Kapitza resistance). Moreover, a specific
relationship between the ratio of heat current and the overlap of
the phonon spectra is demonstrated numerically. The rectifying
effect can also be achieved by modulating the periodicity of the
on-site potential of the FK lattice \cite{Hu05}.

The above works demonstrate the possibility of controlling heat
current by changing structures/parameters of anharmonic lattices.
These might find potential application in energy saving material.
However, almost all works so far are focused on 1D system of
finite size. Obviously, much progress has been achieved, but the
final purpose is to put these ideas to application. It is thus a
nature step forward to seek effective thermal devices to control
heat current in higher dimension. The open questions are: whether
the heat control mechanism in 1D is still valid for the high
dimension(s) and whether the extra dimension(s) reduces the
rectifying efficiency? The answer might not be trivial, as in
higher dimension the lattice vibration includes not only the
longitudinal one but also the transverse ones. The longitudinal
modes will couple to the transverse modes \cite{Wang04}, which
might affect the rectifying efficiency.

In this paper, we concentrate our study on a 2D rectifier model.
We will demonstrate with numerical evidence that a 2D rectifier
can be built up and it works in a very wide temperature range. The
2D thermal rectifier shows similar behaviors with 1D thermal
rectifier under similar parameters regimes.

The paper is organized as the follows. In Section II, we describe
our model and numerical method used for the computer simulation.
In Section III, we demonstrate and discuss the dependence of the
thermal rectifying efficiency on the temperature and the
temperature gradient. Section IV is devoted to the interface
thermal resistance which is the key point to understand the
thermal rectifying effect. In Section V, we give physical
understanding of the rectifying effect in terms of the lattice
vibration spectra (also called phonon band.) We conclude the paper
by conclusions and discussions in Sec. VI.

\section{Model and methodology}

\begin{figure}
\includegraphics[width=\columnwidth]{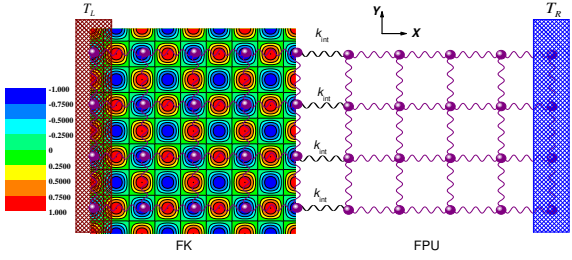}
\vspace{-.8cm}\caption{ (Color Online) Configuration of a two
dimensional (2D) thermal rectifier from the Frenkel Kotoroval and
the Fermi-Pasta-Ulam lattices. The left part is a 2D FK lattice
and the right one is a 2D Fermi-Pasta-Ulam lattice. The two parts
are connected by harmonic springs with constant $k_{int}$. The
left and the right ends are put into contact with heat baths of
temperature $T_L$ and $T_R$, respectively.} \label{Fig:config}
\end{figure}

In our previous work\cite{Li05}, we construct a 1D thermal diode
model by connecting a FK lattice to a FPU lattice with a weak
harmonic spring. We denote it as 1D-FK-FPU model. This model
displays a very good rectifying effect. In this paper, we extend
the 1D- FK-FPU thermal rectifier model to two dimensional one.  We
denote it as a 2D-FK-FPU model. The configuration of the 2D-FK-FPU
model is illustrated in Fig\ref{Fig:config}. The left part is a
plane of harmonic oscillators on a substrate whose interaction is
represented by a sinusoidal on-site potential. Here we plot the
contour line of the 2D sinusoidal potential. For simplicity, we
put one particle in each valley. The right part is an array of
an-harmonic oscillators represented by the FPU model. The two
parts are connected by weak harmonic springs. The Hamiltonian of
the system can be written as:
\begin{equation}
H=H_{FK}+H_{FPU}+H_{int},
\end{equation}
where $H_{FK}$, $H_{FPU}$, $H_{int}$ is the Hamiltonian of the
left part, the right part and the interface section, respectively.
They are represented in (\ref{eq:Hamfk}), (\ref{eq:Hamfpu}),
(\ref{eq:Hamint}), respectively.

\begin{equation}
\begin{split}
H_{FK}&=\sum_{i=1}^{N_{FK}} \sum_{j=1}^{N_Y} ( \frac{\vec
p^2_{i,j}}{2}+k_{FK}V_H(|\vec r_{i,j;i-1,j}|-l_0)\\
&+k_{FK}V_H((|\vec {r}_{i,j;i,j-1}|-l_0))+U_{FK}(x_{i,j},y_{i,j}))
\end{split}
\label{eq:Hamfk}
\end{equation}

\begin{align}
\begin{split}
H_{FPU}&=\sum_{i=N_{FK}+1}^{N_X} \sum_{j=1}^{N_Y} ( \frac{\vec
p^2_{i,j}}{2}\\
&+k_{FPU}V_{FPU}(|\vec r_{i,j;i-1,j}|-l_0)\\
&+k_{FPU}V_{FPU}(|\vec r_{i,j;i,j-1}|-l_0))
\end{split}\label{eq:Hamfpu}
\end{align}
\begin{equation}
H_{int}=\sum_{j=1}^{N_Y}k_{int}V_H(|\vec
r_{N_{FK},j;N_{FK}+1,j}|-l_0) \label{eq:Hamint}
\end{equation}

where $\vec r_{i,j;k,l}=\vec q_{i,j}-\vec q_{k,l}$ is the relative
displacement between particles, labelled as  $(i,j)$ and $(k,l)$.
$V_H(x)=\frac12 x^2$, $V_{FPU}(x)=\frac12 x^2 +\frac14 x^4 $,
$U_{FK}(x,y)=-\frac{A}{(2\pi)^2}\cos(\frac {2\pi}{l_0} x)\cos(\frac
{2\pi}{l_0} y)$.

The mass of the particles is uniformly 1. $l_0$ is distance between
nearest neighbors in equilibrium. The particle at the $i$th column
and the $j$th row is labelled as $(i,j)$. The coordinate and
momentum of this particle is $\vec q_{i,j}=(x_{i,j},y_{i,j})$ and
$\vec p_{i,j}=(p_{x_{i,j}},p_{y_{i,j}})$. In order to establish a
temperature gradient, the two ends of the planes are put into
contact with two Nos\'e-Hoover heat bathes\cite{Nose} with
temperature $T_L$ and $T_R$ for the left end and the right end,
respectively. Particles for $i=1, j=1,2,3...N_Y$ are coupled with
heat bath of temperature $T_L$ and particles for $i=N_X,
j=1,2,3...N_Y$ are coupled with heat bath of temperature $T_R$. We
checked in 1D case that the result does not depend on the particular
heat bath realization. Fixed boundary condition is used along
temperature gradient direction, denoted as $X$ direction in this
paper, namely, $\vec q_{0,j}=(0,j)$, $\vec q_{N_X+1,j}=(N_X+1,j)$.
Periodic boundary condition is applied in the $Y$ direction, namely,
$\vec q_{i,1}=\vec q_{i,N_Y+1}$ (see Fig. 1). Under these boundary
conditions, the system can be considered as a tube. The total number
of particles is $N_X\times N_Y$. All results given in this paper are
obtained by averaging over $N\times 10^8$($N>2$)steps after a
sufficient long transient time when a non-equilibrium stationary
state is set up. The equations of motion of the particles are:
\begin{equation}
\begin{array}{c}
\dot{\vec{q}}_{i,j}=\vec{p}_{i,j},\\
\\
\dot{\vec{p}}_{i,j}=\left\{
          \begin{array}{ll}
          -\frac{\partial H}{\partial \vec {q}_{i,j}} (i=2,N_X-1),\\
\\
          -\frac{\partial H}{\partial \vec {q}_{i,j}}-\xi_{i,j}\vec
          p_{i,j} (i=1,N_X).\\
          \end{array}
          \right.\\
\end{array}
\end{equation}

and the auxiliary variables $\xi_{i,j}$ are described by the
equations:
\begin{equation}
\dot{\xi}_{i,j}=\frac{1}{\mathcal{Q}}(\frac{\vec{p}_{i,j}^2}{2k_BT}-1),
\end{equation}
here, $T$ is the temperature of the heat bath ($T_L$ or $T_R$), and
 $\mathcal {Q}$ is the parameter of coupling between the thermal bath and the
system. In this study, we set $\mathcal {Q}=1$  so that the response
time of the thermostats, $\frac{1}{\sqrt{\mathcal Q}}$, is of the
same order of the original time scale of the lattice. Our purpose is
to study rectifying effect in two dimension and the dependence of
the rectifying efficiency on the system temperature and the
temperature gradient, so we don't attempt to search the optimum
setting of parameters. We choose the system parameters same as in 1D
FK-FPU model which has been tested as good one in 1D case, that is
$k_{FK}=1$, $A=5$ , $l_0=1$, $k_{FPU}=0.2$ and $k_{int}=0.05$. We
set
\begin{equation}
\begin{array}{c}
T_L=T_0(1+\Delta),\\
\\
T_R=T_0(1-\Delta),
\end{array}
\end{equation}
where $-0.8\le\Delta\le 0.8$ in our simulations. So we can simply
denote $T_0$ as the temperature added on the system and
$2\Delta=(T_R-T_L)/T_0$ as the normalized temperature difference of
the system.

The temperature used in our numerical simulation is dimensionless.
It is connected with the true temperature $T_r$ through the
following relation\cite{Hu98}, $ T_r=\frac{m\omega_0^2b^2}{k_B}T,$
where $m$ is the mass of the particle and $b$ is the period of
external potential. $\omega_0$ is the vibration frequency. $k_B$
is the Boltzman constant. For the typical values of atoms, we have
$T_r\sim(10^2-10^3)$\cite{Hu98}, which means that the room
temperature corresponds to the dimensionless temperature $T \sim
(0.1 - 1)$.

The local temperature is defined as
\begin{equation}
T_{i,j}=\frac{1}{2}m \langle\vec {v}_{i,j}^2\rangle,
\end{equation}
where $\langle \dots  \rangle$  stands for a temporal average. The
local heat current $J_{i,j}$ is defined as the energy transfer per
unit time from the particle labelled as $(i,j)$ to the nearest
particles along $X$ direction.
\begin{equation}
\begin{split}
J_{i,j}&=-\vec{v}_{i,j}\cdot \vec{F}_{i,j}\\
&=-k\vec {v}_{i,j}\cdot \frac{\partial\left(V(|\vec
r_{i+1,j;i,j}|-l_0)+V(|\vec r_{i,j;i-1,j}|-l_0) \right)}{\partial
\vec{q}_{i,j}},
\end{split}
\end{equation}
where $k=k_{FK},k_{FPU}$ or $k_{int}$, depending on the site along
$X$ direction. For a 2D-lattice, we treat only heat current flowing
along the $X$-direction. We denote the current from the particles in
$i$th section to the next section in the $X$-direction simply as
$J_i$($J_i=\sum_{j=1}^{N_Y} J_{i,j}$). The total current of the
system is averaged over all sections,
\begin{equation}
J= \frac{1}{N_X}\langle\sum_{i=1}^{N_X}J_i\rangle.
\end{equation}

In our simulations, the fluctuations of temporal heat current
through each section are all less than five percents. We use
$|J_+/J_-|$ as a {\it rectifying efficiency}, to describe
quantitatively the rectifying performance of the system. $J_+$ is
the current when $ \Delta $ is positive (heat flows from the FK
part to the FPU part) and $J_-$ is the current when $\Delta$ is
negative (heat flows from the FPU part to the FK part).

\begin{figure}
\includegraphics[width=\columnwidth]{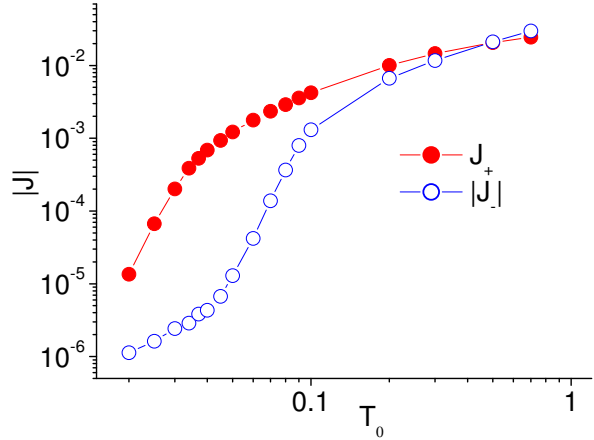}
\vspace{-.8cm}\caption{(Color Online) Heat current of the system
with $N_Y=4$ and $N_X=10$ versus $T_0$. $|\Delta|$=0.5}
\label{fig:rectifying}
\end{figure}

The model we used is a simple extension from one dimension to two
dimension, however the results in Fig\ref{fig:rectifying} show
that it truly demonstrates good rectifying effect on heat current.
In Fig \ref{fig:rectifying}, we can see the visible difference
between $J_+$ and $J_-$ in very wide temperature range. The
difference varies from few times to several hundreds times.

\section{Dependence of rectifying effect on temperature and temperature difference}

\begin{figure}
\includegraphics[width=\columnwidth]{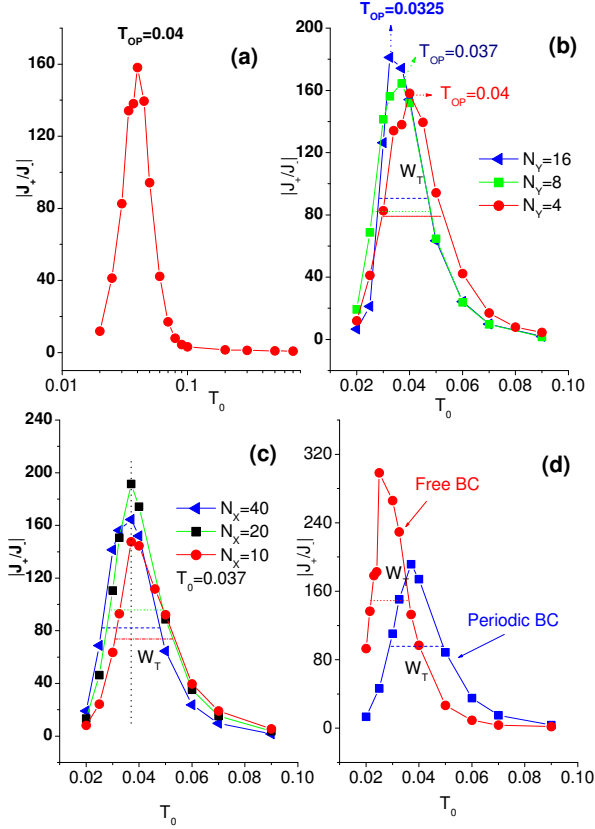}
\vspace{-.8cm}\caption{(Color Online) The rectifying efficiency,
$|J_+/J_-|$, versus $T_0$ at different conditions. (a)The system
with $N_Y=4$, $N_X=10$, and $|\Delta|=0.5$. $T_{OP}=0.04$. (b)
Comparison for three different $N_Y$ with $N_X=40$. $T_{OP}=0.04$,
$W_T=0.022$ and $\vartheta =0.55$ for the case of $N_Y=4$. When
$N_Y=8$, $T_{OP}=0.037$, $W_T=0.022$ and $\vartheta =0.59$. When
$N_Y=16$, $T_{OP}=0.0325$, $W_T=0.018$ and $\vartheta =0.55$.(c)
Comparison for three different $N_X$ with $N_Y=8$. They have the
same $T_{OP}=0.037$. When $N_X=10$, $W_T=0.023$ and $\vartheta
=0.62$. When $N_X=20$, $W_T=0.020$ and $\vartheta =0.54$. When
$N_X=40$, $W_T=0.022$ and $\vartheta =0.59$. (d) Comparison for
different boundary conditions along $Y$ direction with $N_Y=8,
N_X=20$. Free boundary: The optimum performance (OP)$OP=298.5$,
$T_{OP}=0.025$, $W_T=0.014$, $\vartheta =0.56$; Periodic boundary:
$OP=191.5$, $T_{OP}=0.037$, $W_T=0.022$, $\vartheta =0.59$.}
\label{Fig:environment1}
\end{figure}

In this section, we study the dependence of the system performance
on the temperature change. Fig \ref{Fig:environment1} shows the
rectifying efficiency $|J_+/J_-|$ versus $T_0$. In
Fig\ref{Fig:environment1}(a), we can see that there exists an
optimum performance (OP) of the rectifying effect when changing
temperature $T_0$. We can define the temperature for the optimum
performance as $T_{OP}$. In Fig\ref{Fig:environment1}b,
Fig\ref{Fig:environment1}c and Fig\ref{Fig:environment1}d we show
the dependence of the ratio $|J_+/J_-|$ on temperature under
different conditions. We found that $T_{OP}$ depends on the system
settings along $Y$ direction. In Fig\ref{Fig:environment1}b, the
number of particles along $Y$ direction varies from 4 to 8 and 16
while other settings are kept unchanged. We can see clearly that
$T_{OP}$ shifts to lower temperature when increasing $N_Y$. The
value of $T_{OP}$ is 0.04, 0.037, 0.0325 for $N_Y=4$, 8 and 16,
respectively. $T_{OP}$ keeps the same value when we change $N_X$.
This is shown in Fig\ref{Fig:environment1}c. In
Fig\ref{Fig:environment1}d, we change the periodic boundary
condition in $Y$ direction to free boundary condition.  $T_{OP}$
and the optimum performance change drastically. $T_{OP}$ changes
from 0.037 to 0.025 and the ratio increases almost 100\%.

Quantity $W_T$ in the figures is a parameter defined as the width
of the effective temperature range $T_e$ over the half value of
OP, while $\vartheta=W_T/T_{OP}$ is defined as the \emph{quality
factor}. $T_e$, $W_T$, and $\vartheta=W_T/T_{OP}$ are useful
parameters to estimate the temperature range in which the system
has a good rectifying effect. In our investigation, $T_e$ broadens
from the center [0.03-0.46] to high temperature region or low
temperature region. The typical value of $W_T$ under different
settings is around 0.018-0.023. $\vartheta$ is always larger than
0.5. The results suggest that the system is effective in very wide
temperature range.

\begin{figure}
\includegraphics[width=\columnwidth]{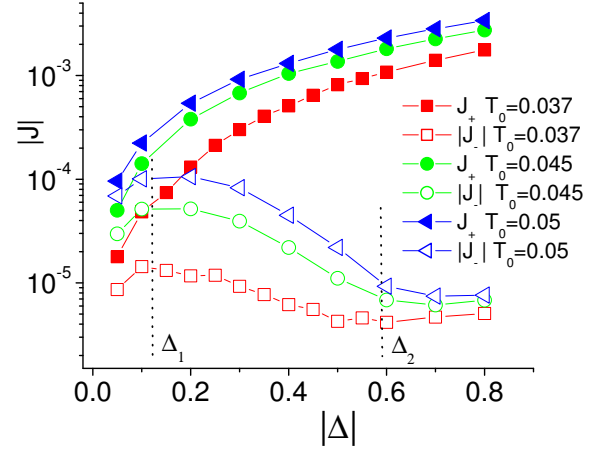}
\vspace{-.8cm}\caption{(Color Online) Heat current of the system
with $N_Y=8, N_X=20$ versus temperature difference at three
different $T_0$. Note that $|J_-|$ decreases as $|\Delta|$
increases in the interval $0.1<|\Delta|<0.6$, this is the
so-called "\emph{negative differential thermal resistance}", see
context for more explanation.} \label{environment2}
\end{figure}

 In Fig\ref{environment2}, we show the heat current versus the (normalized)
 temperature difference, $\Delta$, for three different $T_0$. Full
symbols represent $J_+$ and empty ones $|J_-|$.  We can see that
$J_+$ increases with $\Delta$ monotonically and it always larger
than $|J_-|$, while $J_-$ changes with $\Delta$ in different ways.
One can find that there are three regions for $|J_-|$. In the
first region, $|\Delta|<0.1$, the increase of $\Delta$ leads to
the increase of the heat current. However, in the second region,
$0.1<|\Delta|< 0.6$, the increase of $|\Delta|$ does not induce
the increase of $|J_-|$, instead it results in a decrease of
$|J_-|$. In the third region, $\Delta>0.6$, $|J_-|$ is almost a
constant  independent of $|\Delta|$. In this region, the $|J_-|$
is so small
 that the system can be approximately
considered as an insulator.  The ratio $|J_+/J_-|$ becomes larger
and larger when  $|\Delta|$ increases. That indicates that
rectifying effect increases with increasing temperature
difference.

The strange behavior of $|J_-|$ observed in the second region,
namely, the larger the temperature difference the smaller the heat
current, is called \emph{negative differential thermal
resistance}. As we will demonstrate later that this is a typical
phenomenon in nonlinear lattices. It can be understood from the
match and mismatch of the vibrational spectra of the interface
particles.

\begin{figure}
\includegraphics[width=\columnwidth]{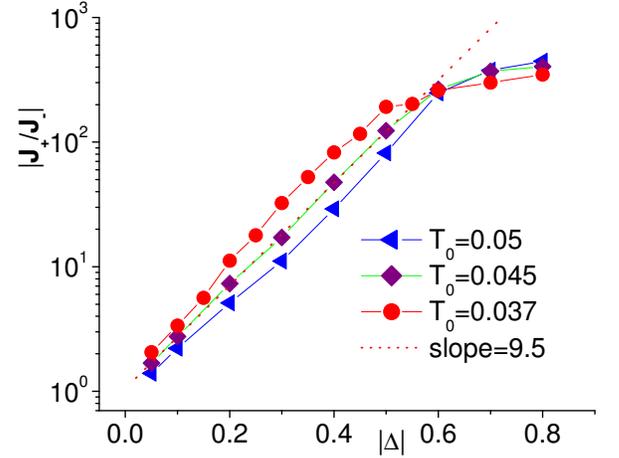}
\vspace{-.8cm}\caption{ (Color Online) The ratio of heat current,
$|J_+/J_-|$ versus half normalized temperature difference
$|\Delta|$ at temperature $T_0=0.037, 0.045, 0.05$. The dotted
line has a slope of 9.5.} \label{Fig:current ratio}
\end{figure}

Fig\ref{Fig:current ratio} shows $|J_+/J_-|$ versus $|\Delta|$. It
is found that $|J_+/J_-|$ increases with $|\Delta|$ in an
exponential way in the regime in which $|\Delta|$ is smaller than
$\Delta_2$. Approximately,
\begin{equation}
|J_+/J_-|\propto \exp(c*|\Delta|)+\delta, \label{eq:approx}
\end{equation}
here $c$ is about 9.5 under the particular parameter setting
$N_Y=8$ and $N_X=20$ with periodic boundary condition along $Y$
direction. When we change $N_Y$, $N_X$ or boundary condition along
$Y$ direction, $c$ changes slightly, but it is always around 10.
The variation of $|c-10|$ is smaller than 0.5 in our
investigation. $\delta$ is related with $T_0$ and $\Delta$.

\begin{figure}
\includegraphics[width=\columnwidth]{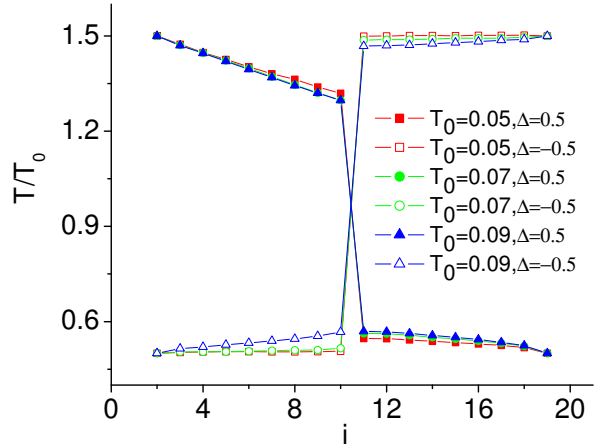}
\vspace{-.8cm}\caption{(Color Online) $T/T_0$ versus lattice site
for $T_0$=0.05, 0.07, and 0.09. The solid symbols are for the
cases of $\Delta =0.5$ and the empty symbols are for the cases of
$\Delta =-0.5$. $N_Y=8, N_X=20$.} \label{Fig:temperature}
\end{figure}

\section{Interface Thermal Resistance (ITR) - Kapitza resistance}

Thermal resistance between two different materials or between twin
or twist boundaries of the same material has been extensively
studied both experimentally and theoretically
\cite{Kinder,Nakayama,Cahill,Kapitza}. In fact, the existence of a
thermal boundary resistance between a solid and superfluid helium
was first detected by Kapitza\cite{Kapitza} as early as 1940's.
This boundary resistance is named Kapitza resistance after him.
Later, it is found that such a Kaptiza resistance exists at the
interface between any pair of dissimilar materials.
Khalatnikov\cite{Khalatnikov} developed the acoustic mismatch
model to explain the Kapitza resistance. Since then, continuous
efforts have been devoted to this problem.
 More information can be found in the review by Swartz and Pohl \cite{SwartzPohl}.

 The Kapitza resistance is defined as
\begin{equation}
R=\Delta T/J, \label{eq:interR}
\end{equation}
where $J$ is the heat current and $\Delta T$  the temperature
difference between two sides of the interface.

In our system, the temperature drops are different when the
temperature gradient of the system is reversed as shown in Fig. 6.
Therefore, we use $R_+$ and $R_-$ to denote the interface resistance
for the case of $\Delta>0$ and $\Delta<0$, respectively.

\begin{figure}
\includegraphics[width=\columnwidth]{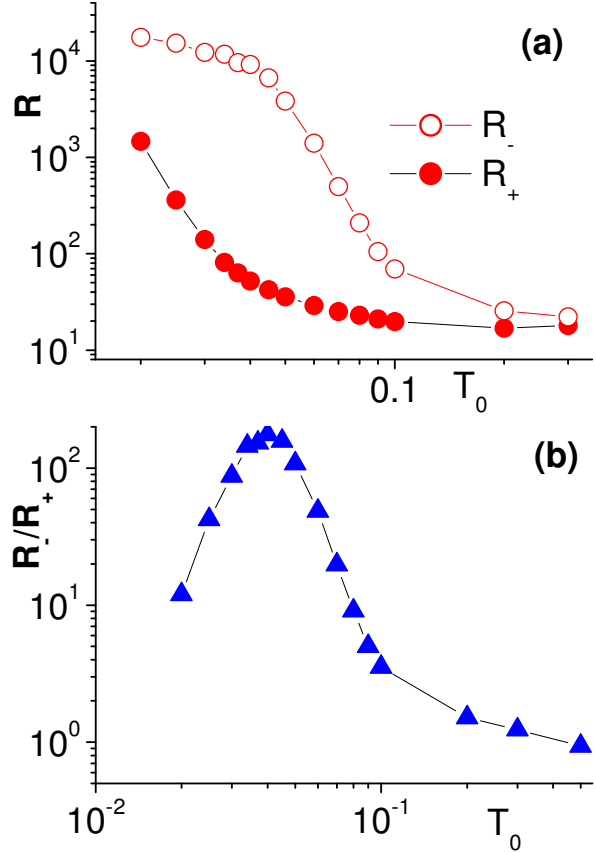}
\vspace{-.6cm}\caption{(Color Online) (a) Interface thermal
resistance $R_{\pm}$ versus $T_0$. $|\Delta|=0.5$. (b) The ratio
$R_-/R_+$ versus $T_0$. $N_Y=4, N_X=10$.} \label{Fig: resistance1}
\end{figure}

In Fig\ref{Fig: resistance1} and Fig\ref{Fig: resistance2}, we show
the dependence of IRT on temperature $T_0$ and the normalized
temperature difference $\Delta$. From the two figures we can see
that, generally, $R_-$ (with larger temperature drop) is about two
or three orders of magnitude larger than $R_+$ (with smaller
temperature jump). Both $R_+$ and $R_-$ decrease with $T_0$ until
$T_0 \approx 0.2$, then both become approximately constants. In
Fig\ref{Fig: resistance1}b, we show the ratio $R_-/R_+$ versus
temperature $T_0$. It is clearly seen that there exists an optimal
temperature value for the ratio $R_-/R_+$. In Fig\ref{Fig:
resistance2}, we show the resistance versus temperature difference.
We can see that $R_-$ monotonically increases with temperature
difference until it reaches a maximum value, while $R_+$
monotonically decreases with increasing $\Delta$. If we plot the
ratio of $R_-$ over $R_+$ versus temperature difference, we find
that the relationship between them also obeys the exponential law
like the ratio of heat current.

\begin{figure}
\includegraphics[width=\columnwidth]{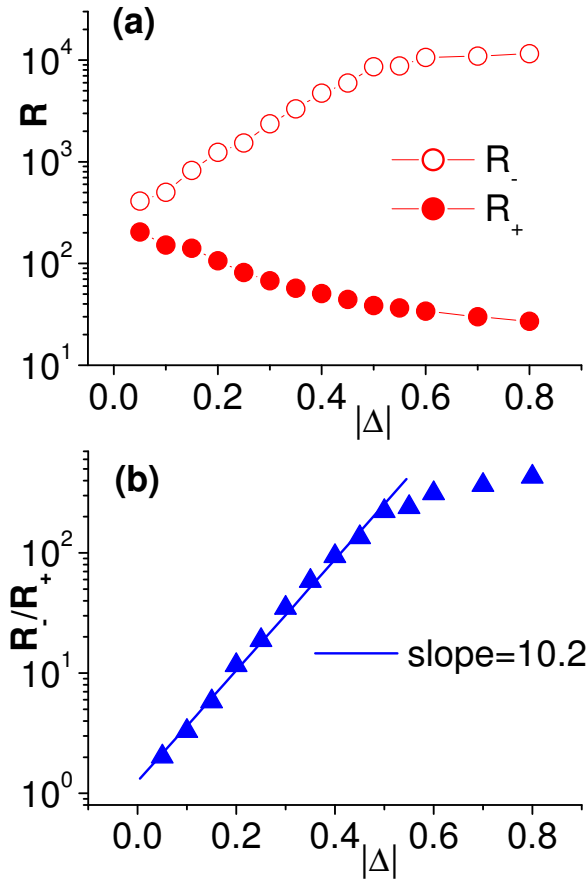}
\vspace{-.6cm}\caption{(Color Online) (a) $R_{\pm}$ versus the
normalized temperature difference $|\Delta|$. $T_0=0.037$. (b) The
ratio $R_-/R_+$ versus $|\Delta|$. $N_Y=8, N_X=20.$} \label{Fig:
resistance2}
\end{figure}

Comparing Fig\ref{Fig: resistance1} and Fig\ref{Fig: resistance2}
with Fig\ref{Fig:environment1} and Fig\ref{Fig:current ratio}, we
can find that the behaviors of the IRT and heat current through
the system is very similar, both are asymmetric, both the ratio
$R_-/R_+$ and $J_+/J_-$ have an optimum value under different
temperature and obey exponential law when changing temperature
difference. The asymmetry of thermal resistance when reversing
temperature gradient is the determinant factor for the rectifying
effect on heat current of the system from the formula
(\ref{eq:interR}).

\section{Physical Mechanism of Rectifying Effect: An Analysis of
Lattice Vibration Spectrum}

From above investigation, we know that the asymmetric interface
thermal resistance determines the asymmetry of heat current when
reversing temperature gradient on the system, but what cause the
asymmetrical behavior of the ITR? In this section, we will answer
this question from a fundamental point of view: lattice vibration
spectrum. Lattice vibration is responsible to heat transport in our
model. An effective way to get the spectrum of lattice thermal
vibration is the discrete faster Fourier transform (DFFT) for the
lattice velocity \cite{William}.

In our model, the vibration has two components, $v_x$ and $v_y$. We
can use the theorem of equipartition of energy to simplify the
numerical calculation. According to the equipartition theorem, the
molecules in thermal equilibrium (here we have local thermal
equilibrium) have the same average energy associated with each
independent degree of freedom of their motion and that energy is
$k_BT/2$. For our system, we have $m\langle
\textbf{v}^2\rangle/2=k_BT$, $m\langle v_x^2\rangle/2=m \langle
v_y^2\rangle/2=k_BT/2$.

\begin{figure}
\includegraphics[width=\columnwidth]{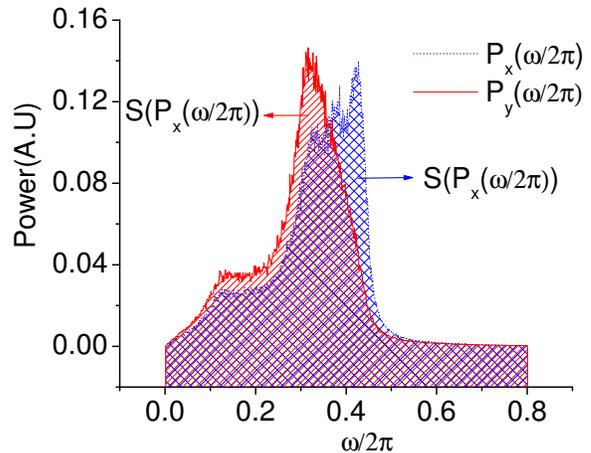}
\vspace{-.8cm}\caption{ (Color Online) DFFT of $v_x, v_y$. The
shadowed regions represent the integral of the power spectra.
$2S(P_x(\omega/2\pi)/L)=0.05=\langle v_x^2\rangle\approx
T$,$2S(P_y(\omega/2\pi)/L)=0.05=\langle v_y^2\rangle\approx T$ . }
\label{Fig:current power}
\end{figure}

In our calculation, $m=1$, $k_B=1$. So we have $\langle
v_x^2\rangle=\langle v_y^2\rangle=T$. If we do the DFFT of $v_x$
and $v_y$ separately, we should have, $
\sum_{i=0}^{L-1}|v_x(t)|^2=\Delta f*\sum_{j=0}^{L/2-1}2|P_x(j)|^2$
and $ \sum_{i=0}^{L-1}|v_y(t)|^2=\Delta
f*\sum_{j=0}^{L/2-1}2|P_y(j)|^2$. Here $P(f)$ is the Fourier
transform of velocity $v(t)$. From the equipartition theorem, one
has $\sum_{i=0}^{L-1}|v_x(t)|^2/L=\langle v_x^2\rangle$ and $
\sum_{i=0}^{L-1}|v_y(t)|^2/L=\langle v_y^2\rangle $ From above
analysis, we have $ \Delta f \sum_{j=0}^{L/2-1}2|P_x(j)|^2/L=T,
\quad \Delta f\sum_{j=0}^{L/2-1}2|P_y(j)|^2/L=T.$ The power
spectra of vibration obtained from DFFT of $v_x$ and $v_y$ agree
with the above formula very well (see Fig\ref{Fig:current power}).
The integral of power spectra is exactly equal to the temporal
average of $v_x^2$ and $v_y^2$ individually. There is a slight
difference between $\langle v_x^2\rangle $ and $\langle
v_y^2\rangle$. The difference might be caused by the number of
sampled data or the different boundary conditions in $X, Y$
direction. In the formula, $L$ should be infinite.

In our system, we find that the asymmetrical ITR and heat current
are strongly related with the overlap of vibration spectra of the
particles at the two sides of the interface. When the vibration
spectra overlap with each other, the system behaves like a thermal
conductor, while the system behaves like a thermal insulator when
the vibration spectra are separated.

Physically, whether an excitation of a given frequency can be
transported through a mechanical system depends on whether the
system has a corresponding eigenfrequency. If the frequency matches,
the energy can easily go through the system, otherwise, the
excitation will be reflected. In our system, the overlap of the two
spectra means that there exists common vibrational frequency in two
parts of the system. The excitation (here as phonon) of such common
frequency can be transported from one part to the another. However,
if the vibrational spectra of two parts are separated, then the
excitation at any part cannot be transported to another part,
because there exists no such corresponding frequency in another
part.

The change from overlap to separation is induced by the different
temperature dependence of the vibration spectra of the two
segments, which is a general feature of any anharmonic lattice.

\begin{figure}
\includegraphics[width=\columnwidth]{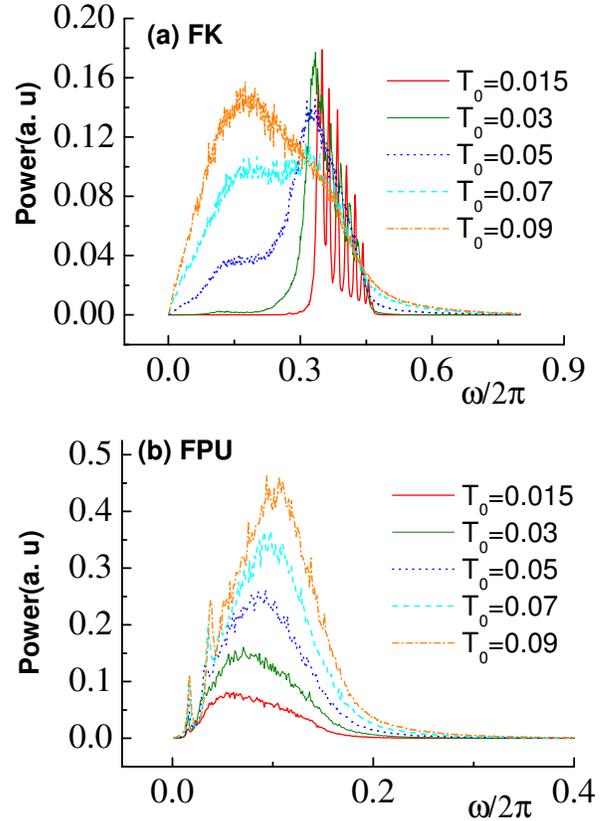}
\vspace{-.6cm}\caption{(Color Online) Vibration spectra of the
interface particles at different temperature. (a) The vibration
spectra of the particles in the FK segment.(b) The vibration
spectra of the particles in the FPU segment.} \label{Fig: spectra}
\end{figure}

In Fig\ref{Fig: spectra}, we show the vibration spectra of the FK
part and the FPU part under different temperatures. We can see
that the vibration spectra of the FK part broaden from high
frequency to low frequency when increasing the system temperature.
This is because at low temperature, the atoms of the FK model is
confined at the valley of the on-site potential, thus the atoms
oscillate in very high frequency, however, when the temperature is
increased, more and more low frequency modes can be excited. In
the limiting case, when the temperature is larger enough that the
kinetic energy of the atom is much larger than the on-site
potential, then the FK model becomes a chain of harmonic
oscillators which has frequency $\omega \in [0, 2\sqrt{k_{FK}}]$.

\begin{figure}
\includegraphics[width=\columnwidth]{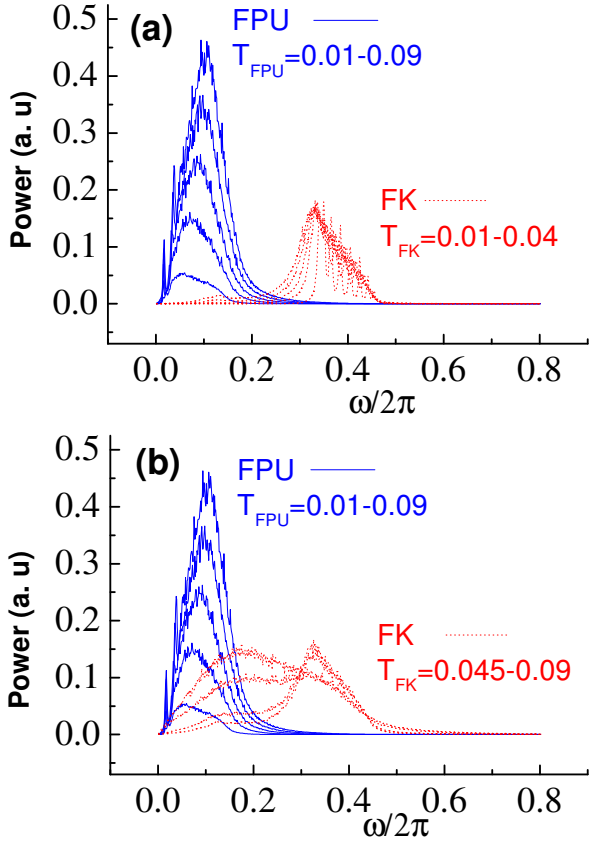}
\vspace{-.8cm}\caption{(Color Online) Comparison of the spectra of
the FK part and the FPU part at different temperature region. (a)
The spectra of the FPU part at $T_{FPU}=0.01-0.09$ (from bottom to
top) and the FK part at $T_{FK}=0.01-0.04$ . (b) The spectra of
the FPU part at $T_{FPU}=0.01-0.09$ and the FK part at
$T_{FK}=0.045-0.09$ } \label{Fig:overlap }
\end{figure}

On the contrary, the vibration spectra of the FPU part broaden from
low frequency to high frequency. In fact, we have shown \cite{Li05}
that the highest oscillation frequency of the FPU model depends on
temperature, $\omega_{FPU} \sim T^{1/4}$. Therefore, in some
settings, the vibration spectra of the FK part and the FPU part will
overlap with each other, while in other temperature settings, they
will separate with each other. These are shown in
Fig\ref{Fig:overlap }. The comparison of vibration spectra of the FK
part at two different temperature ranges and the FPU part at the
full temperature range from 0.01 to 0.12 are shown in
Fig\ref{Fig:overlap }a and Fig\ref{Fig:overlap }b separately. We can
see that the vibration spectra of the FK part and the FPU part are
matched with each other when the temperature of the FK part is from
0.05 to 0.12 (see Fig\ref{Fig:overlap }a). The vibration spectra of
the two parts are separated from each other when the temperature of
the FK part is from 0.01 to 0.03 (see Fig\ref{Fig:overlap }b). This
indicates that when the temperature of the FK part is below a
certain value, which we call $T_c$ ($0.04\sim 0.045$), the system
will behave like a thermal insulator since the separated vibration
spectra of the interface particles make the heat conduction almost
impossible. When the temperature of the FK part is above the
critical point $T_c$, the system will be a good thermal conductor
since the matched vibration spectra allow the heat flow. Thus, if we
adjust $T_0$ and $\Delta$ appropriately to make $T_{high}\geq T_c$
and $T_{low}\leq T_c$ , the system will have a good rectifying
effect. If $T_{high}, T_{low}>T_c$ or $T_{high}, T_{low}<T_c$ , the
rectifying effect is very poor. The above analysis is based on the
vibration spectra of the system with $N_Y=8, N_X=20$. It is
consistent with the result obtained in Section III.

Now we look back at Fig\ref{Fig:environment1}. The optimum point is
at $T_0=0.037$ for $N_Y=8, N_X=20$, corresponding
$T_{high}=0.0505>T_c$ and $T_{low}=0.0235<T_c$. The effective
temperature range $T_e$ with $|\Delta|=0.5$ is from 0.0261 to
0.0481. The low temperatures are all smaller than $T_c$ and all high
temperatures are larger than $T_c$. Both decreasing $T_0$ and
increasing $T_0$ in the outside region of $T_e$ lead the system to
the two extreme cases $T_{high}, T_{low}\geq T_c$ or $T_{high},
T_{low}\leq T_c$ with poor performance. From the above analysis, we
can say that the different properties of heat current under
different temperature and temperature difference are determined by
the temperature dependence of the vibration spectra of the two
segments.

In terms of the vibration spectra of the particles in the interface,
the complex behavior of $|J_-|$ in Fig 4 can be explained from the
vibration spectrum. In particular, in the range from $\Delta_1$ to
$\Delta_2$, a novel phenomenon- called the \emph{negative
differential thermal resistance} phenomenon is observed in Ref.
\cite{BLi04} and fully discussed in Ref \cite{BLi05}. In this
particular temperature interval, a larger temperature difference can
induce a smaller heat current. The \emph{negative differential
thermal resistance} can be understood from the overlap and
separation of the vibration spectra of the interface particles. This
phenomenon is valid for a wide range of the parameters. Moreover, in
Ref \cite{BLi05}, Li. et. al. show that it is this negative
differential thermal resistance property that makes the thermal
transistor possible.

\begin{figure}
\includegraphics[width=\columnwidth]{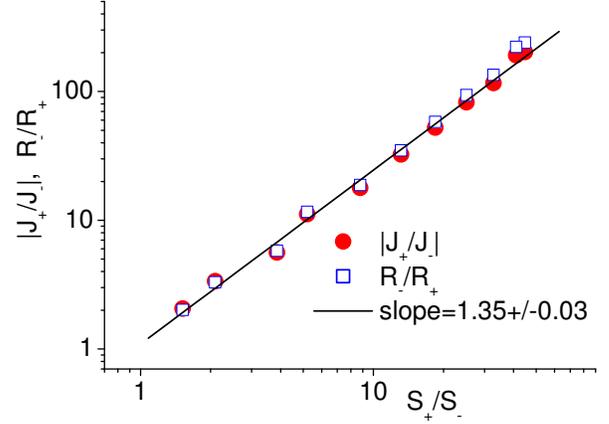}
\vspace{-.7cm}\caption{(Color Online) The bullet are the ratios of
heat current $|J_+/J_-|$ versus $S_+/S_-$ . Square are the ratios
of the ITR $R_-/R_+$ versus $S_+/S_-$. The dotted line has a slope
$1.35\pm0.03$.} \label{Fig: convolution}
\end{figure}

More importantly, we find a specific relationship between the
overlap of the vibration spectra of the two segments and the ratio
$R_-/R_+$ in the interface or the ratio $|J_+/J_-|$ from two
directions.  We introduce the following quantity to describe
overlap of the vibration spectra,
\begin{equation}
S_{\pm}=\frac{\int P^x_l(f)P^x_r(f)df}{\int P^x_l(f)df\int \\
P^x_r(f)df}=\frac{\int P^x_l(f)P^x_r(f)df}{T_{int}^LT_{int}^R}
\end{equation}
$S_{\pm}$ corresponds to the case of $\Delta>0$ and $\Delta<0$,
respectively. In Fig\ref{Fig: convolution}, we plot $S_+/S_-$
versus $R_-/R_+$ and $|J_+/J_-|$. A very good power law was found
between the ratio of resistances or heat currents and the overlap
of the vibration spectra:$
|J_+/J_-|\sim R_-/R_+\propto(S_+/S_-)^\gamma $
The best fit for the ratio of current suggests the power law
constant $\gamma=1.35\pm0.03$. The two figures Fig\ref{Fig:overlap }
and Fig\ref{Fig: convolution} give us a very clear and quantitative
picture about the dependence of the rectifying effect of the system
on the vibration spectra.

\begin{figure}
\includegraphics[width=\columnwidth]{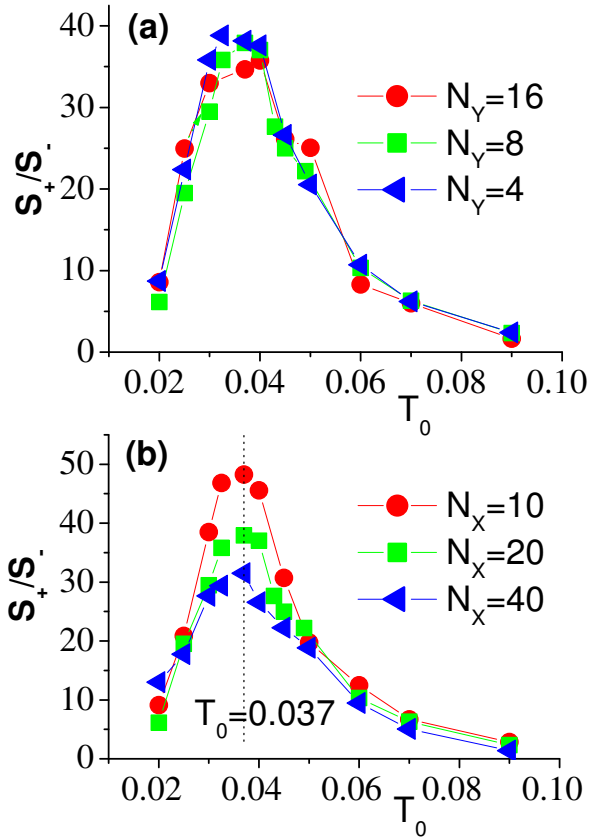}
\vspace{-.6cm}\caption{(Color Online) The ratio of the overlap of
vibration spectra $S_+/S_-$ versus $T_0$ at different conditions.
(a) Comparison at three different $N_Y$ with $N_X=20$.(b)
Comparison at three different $N_X$ with $N_Y=8$. } \label{Fig:
OPspectra}
\end{figure}

Since the rectifying effect sensitively depends on the overlap of
the vibration spectra, we may find answers in Fig\ref{Fig:
OPspectra} for the behavior of the 2D system responding to the
temperature changes at different conditions. We can see clearly
that when we change the number of particles in $Y$ direction, the
temperature for the optimum performance will change, whereas when
we change the number of particles in $X$ direction, the
temperature for OP are kept at the same value. And the value of
the temperature for the OP at different conditions found by the
overlap are consistent with the value in
Fig\ref{Fig:environment1}. These results suggest that the
different vibration spectra of the two segments and the overlap
between them are the determinant factors of the system complex
behaviors.

\section{Discussion and conclusions}

In this paper, we have studied the thermal rectifying effect in a
2D anharmonic lattice. The performance of the 2D thermal rectifier
under different environment changes, such as system temperature
and the temperature difference on the two sides of the system,
have been investigated systematically. We find that there exits an
optimum performance (OP) for a specific thermal rectifier at
certain temperature range. The OP is affected by the boundary
condition and the number of particles, $N_Y$, along $Y$ direction.
The OP shifts to lower temperature when increasing $N_Y$ or
changing the periodic boundary condition to the free boundary
condition along $Y$ direction. The 2D thermal rectifier has a good
rectifying efficiency in a very wide temperature range. Another
important factor that affects the performance of the thermal
rectifier is the temperature difference between the two ends. We
find the rectifying efficiency increases approximately as an
exponential law in certain temperature range with the temperature
difference. The rectifying efficiency is mainly determined by the
asymmetrical ITR. The study on the ITR shows the similar behavior
with heat current.

The behaviors of the ITR and heat current of the system are
strongly correlated with vibration spectra of the particles beside
the interface. The asymmetry behavior of ITR and heat current is
induced by the different temperature dependence of the vibration
spectra of the two parts beside the interface. We find the
vibration spectra of the FK part broaden from high frequency to
low frequency, conversely, the vibration spectra of the FPU part
broaden from low frequency to high frequency as the temperature
increases. The different temperature dependence of vibration
spectra makes the system transition from a thermal conductor to an
insulator possible by setting the system temperature and
temperature difference properly. Moreover a specific relationship
between the performance of the system and the convolution of the
vibration spectra of the two parts is found numerically as power
law.

Our study on 2D thermal rectifier gives a very clear picture about
how the system responds to the environment changes. The results
should be useful for further experimental investigation. The
thermal diode constructed by a monolayer thin film or a tube like
structure might have many practical applications.

\section{Acknowledgement}
We would like to thank Wang Lei for helpful discussions. This work
is supported in part by a FRG of NUS and the DSTA under Project
Agreement No. POD0410553.

\end{document}